# Scalable and Robust Beam Shaping Using Apodized Fish-bone Grating Couplers

**CHAD ROPP,**[1,2] **DHRITI MAURYA,**[1,2] **ALEXANDER YULAEV,**[1,2] **DARON WESTLY,**[1] **GREGORY SIMELGOR,**[1] **AND VLADIMIR AKSYUK**[1*]

[1]*Physical Measurement Laboratory, National Institute of Standards and Technology, Gaithersburg, Maryland 20899, USA*
[2]*Department of Chemistry and Biochemistry, University of Maryland, College Park, MD 20742, USA*
[3]*Physical Measurement Laboratory, National Institute of Standards and Technology, Boulder, Colorado 80305, USA*
*\*vladimir.aksyuk@nist.gov*

**Abstract:**
Efficient power coupling between on-chip guided and free-space optical modes requires precision spatial mode matching with apodized grating couplers. Yet, grating apodizations are often limited by the minimum feature size of the fabrication approach. This is especially challenging when small feature sizes are required to fabricate gratings at short wavelengths or to achieve weakly scattered light for large-area gratings. Here, we demonstrate a fish-bone grating coupler for precision beam shaping and the generation of millimeter-scale beams at 461 nm wavelength. Our design decouples the minimum feature size from the minimum achievable optical scattering strength, allowing smooth turn-on and continuous control of the emission. Our approach is compatible with commercial foundry photolithography and has reduced sensitivity to both the resolution and the variability of the fabrication approach compared to subwavelength meta-gratings, which often require electron beam lithography.

## 1. Introduction

Photonic integrated circuit (PIC) technologies offer a scalable approach to manufacturing photonic systems. PICs often employ grating couplers for converting between on-chip waveguide modes and off-chip modes. These off-chip modes can include optical fiber modes [1], free-space Gaussian beams [2], vortex beams [3], and beams carrying orbital angular momentum states [4]. Coupling to fiber modes requires micron-sized gratings to match the mode size of the fiber, while free-space couplers can range in size from microns to millimeters. Larger grating couplers provide better free-space beam collimation, which is useful for integrated sensors [5], laser cooling [6,7], and light detection and ranging (LIDAR) applications [8].

Efficient power conversion from chip to free-space modes requires apodized grating designs to shape the beam profile and match it with the out-of-plane mode. When designing this apodization, there are tradeoffs constraining the beam shape, the beam size, and the power conversion efficiency [9]. For large gratings, beam shaping is limited by the weakest achievable out-coupling strength, which is typically linked to the minimum feature size set by the lithography approach.

Weaker outcoupling enables larger beam profiles, which can be achieved by fabricating smaller grating elements—typically controlled by either grating depth or width. Reducing the grating depth requires precise control of a partial etch process [6], which is less repeatable than full etch processes and not typically amenable to high-throughput manufacturing. In contrast, reducing the grating widths requires higher resolution lithography. Typical foundry processing can achieve 100 nm to 250 nm resolution using deep UV photolithography. Higher resolution, on the order of 10s of nanometers, can be achieved using electron-beam lithography (EBL), yet this approach is serial and less scalable for manufacturing. EBL also suffers from stitching errors, which can affect the writing of large area grating couplers. Approaches that reduce the grating's effective widths by utilizing subwavelength meta-gratings can be employed to create larger grating couplers at blue wavelengths [9], however, beam shaping was still fundamentally limited by the 30 nm minimum feature size achieved by EBL. Additional solutions can exploit bound-in-continuum modes to decrease the outcoupling strength and provide surface normal emission [10], yet, these structures are resonant and highly narrowband. A recently demonstrated meta-grating approach uses broken symmetry for continuous control of scattering parameters [11], however it requires high-resolution patterning of densely packed elements. Alternative designs are still required to create large area grating couplers compatible with photolithography and commercial foundry processing.

In this article, we develop a grating design based on a fish-bone structure to create apodized millimeter scale beams at 461 nm wavelength. Fish-bone gratings use differential changes to the spine of the grating element to scatter light into free space. These differential changes can be much smaller than the fabrication resolution, which enable weaker scattering and large beam sizes. Fish-bone gratings have previously been developed using partially etched structures for atomic sensing [12], they also have been shown to provide broadband operation in free-standing structures [13] as well as large area beams for LIDAR applications [14,8]. We develop fish-bone gratings with ≈ 100 nm minimum feature sizes and compatibility with fully-etched fabrication processing to create millimeter-scale gratings at 461 nm and 780 nm wavelengths. We tested our designs using EBL and confirmed their performance using a commercial foundry process. Our design combines an evanescent coupler (EVC) [2] with a fish-bone grating to expand the mode profile of a sub-micron sized waveguide into a free-space beam with ≈ 1 mm by 3 mm beam diameter, expanded by the EVC and the grating respectively, thus, achieving greater than a $10^6$ expansion in mode area. The fish-bone grating can achieve both arbitrarily weak scattering strengths and a large apodization dynamic range. We demonstrate fabrication via photolithography of grating elements with amplitudes as low as 2 nm, which is far below the resolution EBL. This amplitude is not fundamental but can be reduced by using a finer resolution apodization design. Additionally, our fish-bone gratings have significantly reduced sensitivity to feature-size dilation caused by fabrication variations compared to subwavelength meta-gratings.

## 2. Design approach

We design our gratings to be compatible with deep-UV photolithography and foundry-scale manufacturing. The device layer consists of a 100 nm thick silicon nitride layer that is fully etched and clad on top and bottom with oxide. This single device layer includes the fish-bone grating, the EVC, routing waveguides, and the spot-size converters used for fiber coupling at the chip facet. We fabricated test devices on an oxidized silicon wafer. The oxide is grown to a nominal thickness of 2.7 µm using a wet thermal oxidation process. The device layer consists of stoichiometric $Si_3N_4$ grown using low-pressure chemical vapor deposition (LPCVD) to a nominal thickness of 100 nm. The EVC and grating elements as well as the fiber couplers at the facet of the chip are patterned using EBL while the waveguides are patterned using a subsequent photolithography step. The fiber couplers include subwavelength structures to expand the waveguide mode to better match that of the fiber [15]. The EVC produces a ≈ 200 µm diameter Gaussian slab-mode beam that is incident normal to the grating [2]. Similar to the gratings, the EVCs can be apodized to generate custom beam profiles that are non-Gaussian. All EBL structures are etched using a single reactive ion etch step through the entire nitride layer thickness. The device layer is then overgrown with a low temperature oxide cladding to a nominal 2.5 µm thickness using LPCVD. The wafer is then diced, and the edges are polished to expose the edge couplers.

The grating design consists of a subwavelength array of identical apodized fish-bone structures, with the spine of the grating element oriented parallel with the propagation direction (Fig. 1a). Figure 1b shows an expanded view of a single fish-bone structure. Each grating "spine" consists of a sinusoidally varying amplitude with a mode-converting taper at the start and the end of the structure. The tapers have a width ($W_{tap}$) defined by the minimum feature size of the lithography approach and a length ($L_{tap}$) that produces an adiabatic transition from the dielectric slab mode to the fish-bone mode. Choosing the lateral grating period, $\Lambda_y$, less than the wavelength of the slab mode prevents lateral scattering in the grating [16] by making it an optical metamaterial. In the design described in this paper, we fix the period of the sinusoid, $\Lambda_x$, and the average spine width $\bar{W}$, however, both parameters can be varied for further apodization control.

By converting from the dielectric slab mode to the fish-bone mode, beam shaping becomes unimpeded by the minimum feature size of the lithography. Zero outcoupling is achieved with zero amplitude modulation of the fish-bone structure. The outcoupling strength can be increased with infinitesimally small increases of the modulation amplitude. This capability contrasts with typical gratings and subwavelength meta-gratings, which are characterized by a finite and discontinuous increase in the outcoupling strength at the beginning of the grating. This discrete increase results from the finite feature size of the smallest grating element. The beam shaping performance of a fish-bone grating is experimentally compared with a subwavelength meta-grating, both operating at 461 nm wavelength (Fig. 1c and 1e, respectively, with profiles of the beams and the apodization design shown in Fig. 1d and 1f).

The fish-bone grating performance is characterized by a smooth transition of the mode profile near the front of the grating, which is indicative of the smooth increase of the outcoupling strength from zero. This contrasts with the meta-grating design, which shows an abrupt transition at the onset of the grating. The apodization of the fish-bone grating is elliptical in amplitude with a constant period, while the meta-grating has linear apodizations of both amplitude and period. The fish-bone apodization is relatively strong, to ensure all power is outcoupled within the 320 µm grating size, which is limited by the write field of the EBL and includes input and output tapers with $T_{tap}$ = 10 µm on either end. Power measurements of the outcoupled light indicate that both designs achieve ≈ 2 dB outcoupling efficiency, with the edge couplers incurring ≈ 5 dB loss [7] and propagation accounting for ≈ 1.3 dB/cm loss.

The fish-bone grating produces a Gaussian-like profile, while the meta-grating can only produce an exponential-like profile, since the minimum outcoupling strength of the meta-grating is too strong to allow shaping of such a large area beam. Both grating designs are fabricated using EBL, a full etch of the device layer, and a subwavelength $\Lambda_y$ period of 250 nm. While the meta-grating has a nominal minimum feature size of 30 nm, the fish-bone grating has a minimum feature size of $W_{tap}$ = 45 nm, but with the minimum line and gap sizes equal to 50 nm and 125 nm, respectively. The smoother fish-bone grating beam profile is achieved using small modulations to the line widths and is not limited by the absolute minimum feature size.

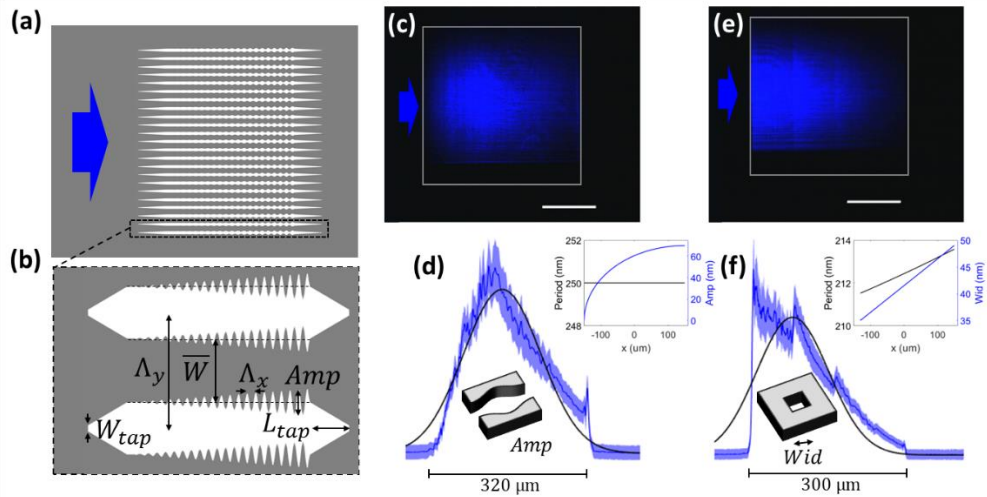

Fig. 1. (a) Fish-bone grating design with slab-mode light incident from the left (blue arrow). Etched regions are shown in white. (b) An enlarged depiction of the beginning and end of the fish-bone grating, consisting of input (left) and output (right) tapers and an apodized sinusoidal structure in the middle. X and Y dimensions are not shown to scale. (c) Image of the light emitted from a fabricated fish-bone grating with a 45 nm minimum feature size and light incident from the left (blue arrow). Grating area outline in gray. Scale bars are 100 µm. (d) Profile of the fish-bone-grating beam with a Gaussian fit (black). The light blue region corresponds to a one standard deviation measurement uncertainty of the beam intensity profile, obtained from different horizontal slices of the profile from panel c. The dark blue line corresponds to the mean profile. Bottom inset shows a fish-bone grating unit cell. Top-right inset shows the period and amplitude of apodization. (e) Image of the light emitted from a fabricated meta-grating with a 35 nm minimum feature size and light incident from the left (blue arrow). Grating outline in gray. (f) Mean profile of the meta-grating beam (blue) with a Gaussian fit (black) and one-standard deviation measurement uncertainty (light blue region). Bottom inset shows a meta-grating unit cell. Top-right inset shows the period and width of apodization.

Figure 2 shows results from fish-bone structures written with both the minimum line and gap sizes ≥ 90 nm, including $W_{tap}$. This feature size is more compatible with the limits of foundry-level photolithography. Devices operating at 461 nm and 780 nm are shown. Despite the much larger minimum feature size, Gaussian beam profiles can still be readily achieved at 461 nm. Instead of limiting beam-shaping, the larger feature size only causes an increase in the scatter at the slab-to-fish-bone mode transition (seen as a small sharp peak on the left in the profile in Fig. 2b). Light scattered here radiates in all directions, rapidly decreasing in intensity in the far-field. Designs at 461 nm and 780 nm are also both well collimated over millimeter length scales (Fig 2g), with a slight divergence and convergence, respectively, that can be compensated with slight apodization of the grating period to flatten the phase front in future designs.

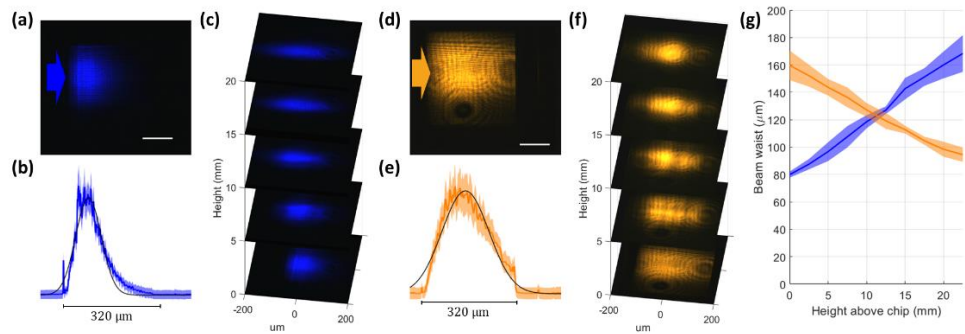

Fig. 2. Beam profiles from fish-bone gratings with a 90 nm minimum feature size. (a) Top-down view of the 461 nm emission from a fish-bone grating with light incident from the left (blue arrow). Scale bars are 100 µm. (b) Beam profile, mean in blue and one standard deviation measurement uncertainty in light blue, with a Gaussian fit (black). (c) Beam imaged at increasing height from the chip surface. (d) Top-down view of the 780 nm emission from a fish-bone grating with light incident from the left (orange arrow). (e) Beam profile, mean in orange and one standard deviation measurement uncertainty in light orange, with a Gaussian fit (black). (f) Beam imaged at increasing height from the chip surface. (g) Fitted beam waist in the x direction at increasing heights above the chip for the 461 nm (blue) and 780 nm (orange) designs. The one standard deviation measurement uncertainty of the 461 nm and 780 nm beam Gaussian fit widths are depicted in light blue and orange, respectively.

## 3. Simulated performance

Eigenmode analysis of the fish-bone grating reveals several advantages compared with line gratings and meta-gratings. Figure 3 plots the calculated real (Figure 3a) and imaginary (Figure 3b) eigenfrequencies as a function of the size parameter that controls the scattering strength. To understand the dispersive behavior of a conventional line grating, we simulate fully etched transverse line gratings with the same 100 nm nitride thickness and varied line widths (blue data). These etched features are much smaller than what typical lithography can achieve. The scattering strength of a line grating is proportional to the imaginary eigenfrequency, which increases quadratically with increasing line width. The simulated trend deviates from quadratic when the grating element is no longer perturbative, with grating line widths greater than $\approx 35$ nm. The change in the real component of the eigenfrequency (Figure 3a) with feature size results from the changing effective index of the unit cell. Such dispersion must be compensated by varying the period of the grating when apodizing the scattering strength.

Compared with line gratings and meta-gratings (gray data), fish-bone gratings (black data) achieve weaker scattering strengths, a larger scattering dynamic range, and lower mode dispersion. Like line gratings, fish-bone gratings scale quadratically with feature size, while meta-gratings scale cubically. This cubic scaling indicates that meta-gratings are more sensitive to fabrication variations than the other two grating designs. Meta-gratings are also the most dispersive of these designs, requiring the most compensation to achieve a flat phase front with an apodized design. While meta-gratings can achieve scattering strengths as weak as sub-resolution line gratings [9], fish-bone gratings can achieve even weaker scattering, down to zero with a 0 nm amplitude.

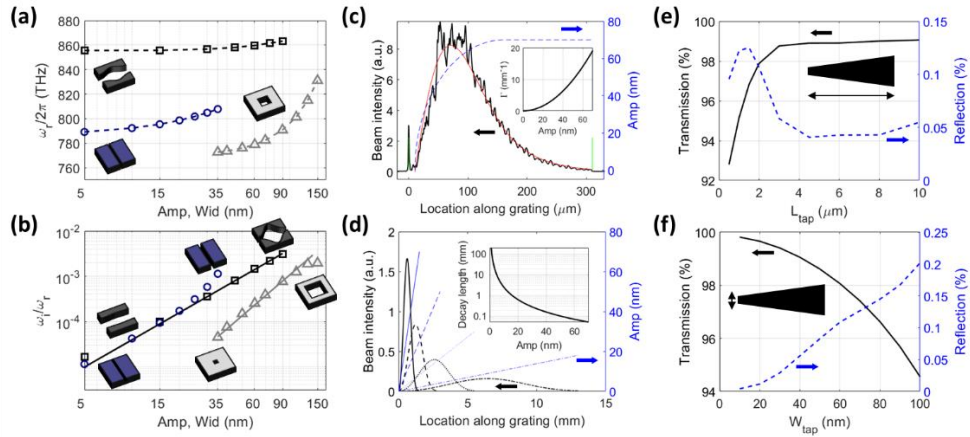

Fig. 3. Performance scaling of fish-bone gratings with feature size. (a) Calculated eigenmode frequency of line gratings (blue circles), meta-gratings (gray triangles), and fish-bone gratings (black squares) with increasing feature size. The dielectric slab is set to be 100 nm thick and the grating period $\Lambda_x = 250$ nm in all cases. (b) Normalized imaginary frequency change with feature size. At small feature sizes both the Bragg and fish-bone gratings scale quadratically with feature size (fit with $2.02 \pm 0.07$ and $2.00 \pm 0.03$ power law trends, respectively), while meta-gratings scale

cubically (fit to a 2.97 ± 0.08 power law). (c) Fitting of an experimentally obtained beam profile (black, with fit in red) with the apodization design (blue, dashed) to empirically fit the correlation between the feature size amplitude to the outcoupling-strength (inset). The green vertical line indicate the extent of the grating used to fit the spatial scale. (d) Simulated beam profiles (black) obtained by linear apodizations (blue lines) using the empirical out-coupling strength correlation fit from panel c and replotted in the inset in units of the decay length. (e) Simulated transmission (black solid) and reflection (blue dash) performance of the insertion tapers as a function of taper length with a constant taper width of 40 nm. (f) Simulated performance as a function of taper tip width with a constant taper length of 10 µm.

Analysis of fabricated devices demonstrate how the weak scattering strength of fish-bone gratings can enable millimeter-scale beams. Figure 3c shows the measured beam profile from a fish-bone grating (same device as in Fig. 2a-c) overlaid with the designed amplitude apodization (blue curve). Assuming a quadratic relationship between the outcoupling strength and the apodization amplitude, we calculate a best fit to the measured data (red curve). The inverse of the outcoupling strength is the decay length. Figure 3d shows how weak linear apodizations can achieve beam waists greater than 1 mm. This data set uses the scattering relationship obtained from the measured data and we model the outcoupled beam profiles (black lines) for different linear apodizations (blue lines) with slopes of approximately {47, 17, 5, 1} nm/mm producing beams with full waists of {1, 2, 4, 10} mm (from left to right).

By allowing infinitesimal perturbation to the initial translation symmetry, the fish-bone gratings enable arbitrarily weak scattering strengths, independent of the minimum feature size. Instead of affecting the beam-shaping performance, the minimum feature size affects the insertion loss of the grating. Scattering at the transition from the dielectric slab mode to the fish-bone mode is determined by the design of the taper (Fig. 1b). Figure 3e and Figure 3f show simulations of how the taper length and the minimum feature size affect the power performance of this transition. With sufficiently long tapers, the transition becomes adiabatic. We calculate that losses can be less than ≈ 5 % when minimum feature sizes are less than 100 nm. Most of these losses are due to scattering, with < 1 % due to reflection.

## 4. Foundry-Fabricated Grating Emitters

While previous grating designs had been limited to ≈ 300 µm by the write field of the EBL, our fish-bone gratings can be made much larger using photolithography, using a process that is also compatible with mass manufacturing. We now fabricate larger gratings using a custom tape-out with a commercial foundry offering 193 nm wavelength photolithography. Similar to the previous fish-bone chips, these chips include spot-size converting edge couplers, waveguides, EVCs, and the fish-bone gratings, which are fabricated using a fully etched 100 nm nitride device layer, which is now grown using plasma-enhanced chemical vapor deposition (PECVD). In our largest device, the EVC creates slab-mode beams with a Gaussian profile and a beam diameter of ≈ 1.2 mm, while the EVC of smaller devices—designed with dimensions to match the previous EBL designs—produce Gaussian beams with diameters of ≈ 200 µm.

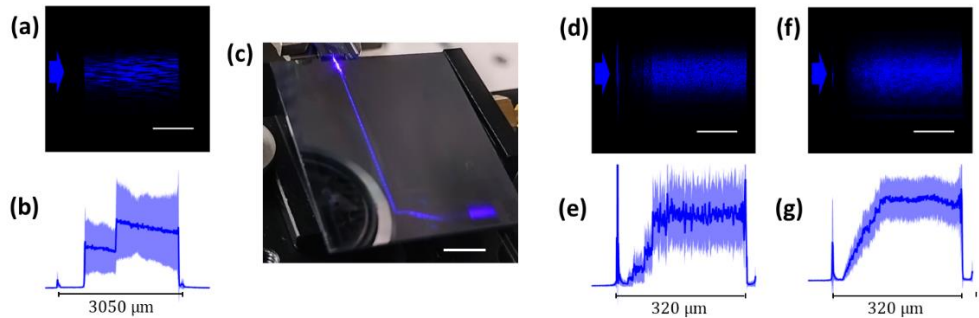

Fig. 4. Apodized beam profiles from foundry-fabricated grating devices. Optical image (a) and beam profiles (b) of a 3 mm long fish-bone grating with a 100 nm minimum feature size and $\Lambda_y$ = 250 nm. (c) Image of the chip showing scattered 461 nm light from an inverse-taper fiber coupler, top left, a connecting waveguide, an ≈ 7 mm long EVC, and a 3 mm long grating device, bottom right. Optical image (d) and beam profile (e) of a 300 µm long apodised fish-

bone grating with a 110 nm minimum feature size and $\Lambda_y$ = 250 nm. Optical image (f) and beam profile (g) of a fish-bone grating with a 120 nm minimum feature size, but $\Lambda_y$ = 550 nm. The scale bar is 1 mm in panel a, 5 mm in panel c, and 100 µm in panels d and f.

Figure 4a-c shows the beam produced by a 3 mm fish-bone grating outcoupler. This grating has apodization with just two steps in the amplitude, from 0 nm to 2 nm, and from 2 nm to 4 nm amplitude. Fig. 4b shows this step-wise beam profile, which has a size much greater than 1 mm. By writing the discrete steps we can distinguish the decay lengths of the three fish-bone amplitudes. The decay lengths are too long to be measured with shorter gratings or accurately estimated from simulation. The 0 nm amplitude has near-zero scattering, while the 2 nm and 4 nm amplitudes produce decay lengths of $(8.0 \pm 0.5)$ mm and $(7.7 \pm 0.1)$ mm, respectively. These decay lengths are very similar to one another, implying that the scattering strength does not follow the expected quadratic dependence. We believe that the amplitude of the fabricated sinusoid is not scaling linearly with the nominal amplitude of the design due to an effect of the mask fabrication or photolithography. Fabricating gratings with a more Gaussian-like profile will require adjusting the apodization to account for this nonlinearity as well as increasing the resolution of our grating design to include effective step sizes finer than 2 nm, such as by using a non-sinusoidal nominal profile shape in the propagation direction.

In addition to the 3 mm grating device, we fabricated devices with different minimum feature sizes, equal to (90, 100, 110, or 120) nm. We found that larger minimum feature sizes gave better beam profiles. Devices with features as small as 90 nm or 100 nm (like is the case for the 3 mm grating, see Fig. 4a) often had fabrication irregularities that produced phase-front distortions and stripes in the beam profile along the direction of the fish-bone spine. Larger feature size gratings produced cleaner beam profiles, but they also have a narrower apodization range. As seen in Fig. 4 d-e, a grating with a 110 nm minimum feature size only has an amplitude range of 10 nm, corresponding to five steps of 2 nm amplitude. The minimum feature size and the lateral period of $\Lambda_y$ = 250 nm physically limit this maximum apodization amplitude. With this weak amplitude, we see that not all the power in the beam is scattered into free-space by the end of the 300 µm grating length. The larger period device shown in Fig. 4f-g, with $\Lambda_y$ = 550 nm, enables a maximum apodization amplitude of 40 nm. The larger $\Lambda_y$ provides a larger range for apodizing the amplitude of the grating at the expense of scattering higher order beams into free space. Still, this larger amplitude is insufficient for outcoupling all the power by the end of the grating.

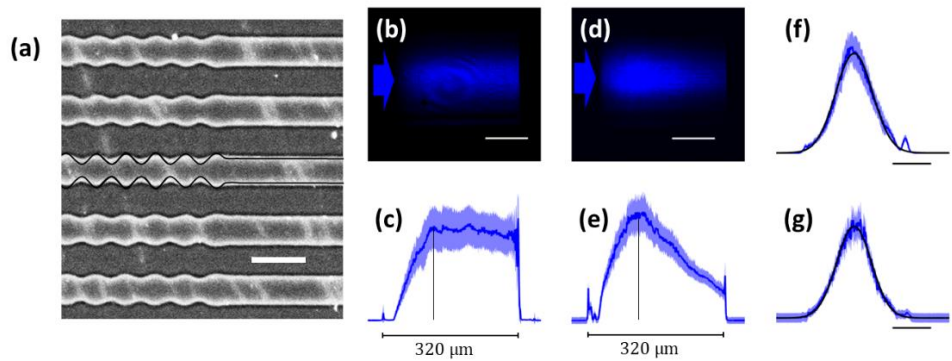

Fig. 5. (a) SEM of the grating written by photolithography showing weaker amplitude modulation than nominally designed. The overlay (black line) depicts an outline of the desired geometry, which was implemented in the mask layout without any proximity effect correction for this test. Scale bar is 500 nm. Apodized beam profiles comparing identical designs fabricated by the foundry's photolithography patterning (b-c) and by EBL (d-e). These devices both have a minimum feature size of 120 nm and $\Lambda_y$ = 550 nm. Despite similar locations of the turn-over points in the beam profiles, the decay tails are much longer for the devices written by photolithography due to the optical lithography producing shallower grating modulation. The lateral profile of the beams are Gaussian-shaped (fit shown

in black) with beam diameters of 191.8 ± 0.4 µm for the device produced by photolithography (f) and 181.3 ± 0.4 µm for the device produced by EBL (g).

To investigate the cause of the weaker scattering strength of the photolithography fabricated gratings we compare them with designs fabricated by EBL. Figure 5a shows an SEM of a 300 µm grating device prepared by photolithography, which has a period of $\Lambda_y$ = 550 nm. While the designed amplitude was drawn to be 100 nm, the amplitude measured from the SEM is only 30 nm. Also, the nitride line width is measured to be ≈ 270 nm, compared to the ideal 220 nm width. These differences likely result from the dense patterning of the grating structure and the fact that the exposure dose was optimized for isolated features, while no correction was implemented on the mask for this test. Future fabrication runs could be improved by adjusting the pattern and the dose. Figures 5b-c and 5d-e compare the profile of beams generated by gratings fabricated by photolithography and EBL, respectively, with the same PECVD nitride device layer. The EBL device has a considerably stronger out-coupling strength, which outcouples the majority of the light energy within the 300 µm grating length. The same exact design written by photolithography has only a weakly decaying tail to the beam shape. Yet, despite the ≈ 70 % difference in the maximum apodization amplitude that appears at the tail end of the grating, the left (input) side of both gratings produce beam profiles that are remarkably similar. In fact, the elbow in the beam profiles occur at roughly the same location (black lines in Fig. 5c and 5e). The lateral profiles of the beams produced by the two devices are also remarkably similar, indicating that the EVC devices came out as expected using photolithography (Fig. 5f and 5g). The EVCs are made up of isolated line features.

## 5. Sensitivity Analysis

As we saw in Fig. 5, the fish-bone grating design offers unique robustness to fabrication variability, producing similar beam profiles even with ≈ 70 % changes to feature sizes. This is in contrast with meta-gratings, which have a strong dependance on feature sizes. Here, we numerically compare how variations to the device layer thickness ($\Delta t$) and the feature etch widths ($\Delta w$) alter the position of the emitted beam along the grating ($\Delta x$) and the emission angle ($\Delta \theta$) (Table 1) for fish-bone and meta-gratings. To better compare the sensitivities, we calculate the position offset at a height T of 500 µm above the grating caused by the angular deviation, denoted (T$\Delta \theta$) as shown in Figure 6a. This offset, caused by an angular deviation, is important when considering the effect of misalignment between the emitted beam and a planar optic located above the chip [7]. For both fish-bone and meta-grating designs, increasing the width of the etched features causes displacements of the beam in the negative direction, while increasing the device layer thickness causes negative displacements, but positive angular deflections.

The most striking comparison in Table 1 is the reduction in $\Delta x/\Delta w$ sensitivity (greater than an order of magnitude) when switching from a meta-grating design to a fish-bone grating design. The effects are demonstrated in Figure 6b-c, which show the simulated beam profile of a 115 µm large Gaussian beam produced by a fish-bone grating compared to a meta-grating, when the etch width is increased by 5 nm (insets show an exaggeration of the feature size increase). While the meta-grating experiences an ≈ 15 µm shift, the fish-bone grating only experiences an ≈ 1 µm shift. The difference can be understood from the inset pictures. When the meta-grating elements increase in size, the out-coupling strength dramatically increases, causing the beam to emit sooner in the grating. In contrast, increasing the etched feature size of the fish-bone grating minimally affects the scattering strength. Instead, the increase will cause more scattering only at the input taper, which does not affect beam shaping performance.

The fish-bone grating also has improved robustness to thickness variations, largely due to the lower index of the fish-bone mode compared to the slab mode of the meta-grating. Spatial sensitivity ($\Delta x/\Delta t$) is better by more than a factor of two, while angular sensitivity (T$\Delta \theta/\Delta t$) is approximately the same. However, not all sensitivities are improved. The angular sensitivity to width variations (T∆θ/∆w) is almost a factor of two worse for fish-bone gratings compared to meta-gratings. This worse response occurs because an increase in the width of the etched region results in more dielectric material removed and a larger change in the effective index of the propagating mode and more deflection angle. So while the fish-bone grating design is drastically more robust to spatial variations, this improvement increases angular sensitivity, which is often less concerning than spatial variations [7].

Table 1. Sensitivity analysis comparing fish-bone gratings and meta-gratings (for T = 500 µm)

| Sensitivity (µm/nm) | ∆x/∆w | T∆θ/∆w | ∆x/∆t | T∆θ/∆t |
| --- | --- | --- | --- | --- |
| Meta-grating | -3.03 | -0.32 | -0.04 | +0.71 |
| Fish-bone grating | -0.14 | -0.57 | -0.02 | +0.68 |

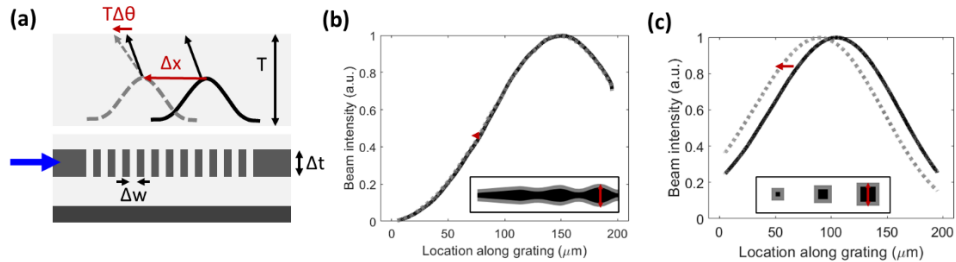

Fig. 6. Beam misalignment sensitivity to fabrication parameters. (a) Diagram showing how changes to the width of the etched feature sizes (∆w) and the thickness of a grating device layer (∆t) can cause beam displacement (∆x) and angular offsets (∆θ) that cause displacements (T∆θ). For this analysis, T is given by 500 µm. The ideal beam location and angle are shown in black, while the offsets are dashed gray. (b) Beam profile of a fish-bone grating design (black) compared to a design with feature sizes increased by 5 nm (gray, dashed). The insets depict the geometry of feature size increase, but with an exaggerated 40 nm increase in size. (c) Beam profile of a meta-grating design (black) compared to a design with feature sizes increased by 5 nm (gray, dashed). The inset depicts the geometry with a 40 nm increase in feature size.

## 6. Conclusions and outlook

By using a fish-bone grating design we have shown that optical scattering can be decoupled from the feature size of the written structure, and that arbitrarily large grating couplers—even at blue wavelengths—may be fabricated with minimum feature sizes compatible with photolithography. Our approach allows for continuous control of the emitted phase profile and the local scattering intensity, with no limit to the minimum scattering strength. We demonstrate scattering strengths down to zero with amplitude step sizes much smaller than can be achieved by EBL in a conventional grating design. In addition, the relatively large range of scattering strength achievable with fish-bone gratings enables apodization of gratings that produce smooth Gaussian beam profiles that would be otherwise impossible with other grating designs. We further demonstrate how fish-bone gratings can create beams with dimensions much larger than 1 mm using commercial deep UV photolithography. Not only are these grating designs more scalable for mass manufacturing, they also prove to be more robust to fabrication variance, especially to feature-size dilation.

This work exclusively studied constant-period fish-bone gratings, which produced nearly-collimated beams, however, by modulating this period, holograms with arbitrary phase and local intensity could be achieved. In addition, this work focused on symmetrically designed fish-bones, yet, future work may explore using a broken fish-bone mirror symmetry to achieve polarization control, similar to the approach taken in Ref. [11]. Better outcoupling power performance may also be achieved by breaking the up-down mirror symmetry of the present structures. By using grating periods near the Bragg condition, forward and backward propagating light can interfere and create resonant and non-local coupling effects [10]. In contrast to the discrete translation symmetry breaking exploited in Ref [11], here we outcouple by breaking a continuous translation symmetry. Our approach may enable efficient and general amplitude, phase, and polarization modulation control without subwavelength patterning of discrete meta-structures.

## Acknowledgements

Dr. Chad Ropp, Dr. Dhriti Mauria and Dr. Alex Yulaev acknowledge support under the Professional Research Experience Program (PREP), administered through the Department of Chemistry and Biochemistry, UMD. Research performed in part at the NIST Center for Nanoscale Science and Technology.

## References

1. X. Chen, C. Li, C. K. Y. Fung, S. M. G. Lo, and H. K. Tsang, "Apodized Waveguide Grating Couplers for Efficient Coupling to Optical Fibers," IEEE Photonics Technology Letters **22**, 1156–1158 (2010).
2. S. Kim, D. A. Westly, B. J. Roxworthy, Q. Li, A. Yulaev, K. Srinivasan, and V. A. Aksyuk, "Photonic waveguide to free-space Gaussian beam extreme mode converter," Light Sci Appl **7**, 72 (2018).
3. A. Liu, C.-L. Zou, X. Ren, Q. Wang, and G.-C. Guo, "On-chip generation and control of the vortex beam," Appl. Phys. Lett. **108**, 181103 (2016).
4. T. Su, R. P. Scott, S. S. Djordjevic, N. K. Fontaine, D. J. Geisler, X. Cai, and S. J. B. Yoo, "Demonstration of free space coherent optical communication using integrated silicon photonic orbital angular momentum devices," Opt. Express, OE **20**, 9396–9402 (2012).
5. M. T. Hummon, S. Kang, D. G. Bopp, Q. Li, D. A. Westly, S. Kim, C. Fredrick, S. A. Diddams, K. Srinivasan, V. Aksyuk, and J. E. Kitching, "Photonic chip for laser stabilization to an atomic vapor with 10^-11 instability," Optica, OPTICA **5**, 443–449 (2018).
6. A. Isichenko, N. Chauhan, D. Bose, J. Wang, P. D. Kunz, and D. J. Blumenthal, "Photonic integrated beam delivery in a rubidium 3D magneto-optical trap," (2022).
7. C. Ropp, W. Zhu, A. Yulaev, D. Westly, G. Simelgor, A. Rakholia, W. Lunden, D. Sheredy, M. M. Boyd, S. Papp, A. Agrawal, and V. Aksyuk, "Integrating planar photonics for multi-beam generation and atomic clock packaging on chip," Light Sci Appl **12**, 83 (2023).
8. W. Xie, T. Komljenovic, J. Huang, M. Tran, M. Davenport, A. Torres, P. Pintus, and J. Bowers, "Heterogeneous silicon photonics sensing for autonomous cars [Invited]," Opt. Express, OE **27**, 3642–3663 (2019).
9. C. Ropp, A. Yulaev, D. Westly, G. Simelgor, and V. Aksyuk, "Meta-grating outcouplers for optimized beam shaping in the visible," Opt. Express, OE **29**, 14789–14798 (2021).
10. A. Yulaev, S. Kim, Q. Li, D. A. Westly, B. J. Roxworthy, K. Srinivasan, and V. A. Aksyuk, "Exceptional points in lossy media lead to deep polynomial wave penetration with spatially uniform power loss," Nat. Nanotechnol. **17**, 583–589 (2022).
11. H. Huang, A. C. Overvig, Y. Xu, S. C. Malek, C.-C. Tsai, A. Alù, and N. Yu, "Leaky-wave metasurfaces for integrated photonics," Nat. Nanotechnol. **18**, 580–588 (2023).
12. M. Puckett, M. Robbins, R. Compton, N. Solmeyer, C. Hoyt, C. Fertig, and K. Nelson, "Integrated photonics for atomic sensing," in *Optical and Quantum Sensing and Precision Metrology* (SPIE, 2021), Vol. 11700, pp. 218–224.
13. Z. Cheng, X. Chen, C. Y. Wong, K. Xu, and H. K. Tsang, "Broadband focusing grating couplers for suspended-membrane waveguides," Opt. Lett., OL **37**, 5181–5183 (2012).
14. W. Xie, J. Huang, T. Komljenovic, L. Coldren, and J. Bowers, "Diffraction limited centimeter scale radiator: metasurface grating antenna for phased array LiDAR," (2018).
15. P. Cheben, P. J. Bock, J. H. Schmid, J. Lapointe, S. Janz, D.-X. Xu, A. Densmore, A. Delâge, B. Lamontagne, and T. J. Hall, "Refractive index engineering with subwavelength gratings for efficient microphotonic couplers and planar waveguide multiplexers," Opt Lett **35**, 2526–2528 (2010).
16. R. Halir, P. Cheben, S. Janz, D.-X. Xu, Í. Molina-Fernández, and J. G. Wangüemert-Pérez, "Waveguide grating coupler with subwavelength microstructures," Opt. Lett., OL **34**, 1408–1410 (2009).